# Redox Potential Replica Exchange Molecular Dynamics at Constant pH in AMBER: Implementation and Validation

*Vinícius Wilian D. Cruzeiro [1,2], Marcos S. Amaral [3], Adrian E. Roitberg [1,*]*

[1] Department of Chemistry, University of Florida, Gainesville, FL 32611, United States

[2] CAPES Foundation, Ministry of Education of Brazil, Brasília – DF 70040-020, Brazil

[3] Institute of Physics, Federal University of Mato Grosso do Sul, Campo Grande, MS 79070-900, Brazil

# ABSTRACT

Redox processes are important in chemistry, with applications in biomedicine, chemical analysis, among others. As many redox experiments are also performed at a fixed value of pH, having an efficient computational method to support experimental measures at both constant redox potential and pH is very important. Such computational techniques have the potential to validate experimental observations performed under these conditions and to provide additional information unachievable experimentally such as an atomic level description of macroscopic measures. We present the implementation of discrete redox and protonation states methods for constant redox potential Molecular Dynamics (CEMD), for coupled constant pH and constant redox potential MD (C(pH,E)MD), and for Replica Exchange MD along the redox potential dimension (E-REMD) on AMBER. Validation results are presented for a small system that contains a single heme group: N-acetylmicroperoxidase-8 (NAcMP8) axially connected to a histidine peptide. The methods implemented allow one to make standard redox potential ($E^o$) predictions with the same easiness and accuracy as $pK_a$ predictions using the CpHMD and pH-REMD methods currently available on AMBER. In our simulations we can correctly describe, in agreement also with theoretical predictions, the following behaviors: when a redox-active group is reduced the $pK_a$ of a near pH-active group increases because it becomes easier for a proton to be attached; equivalently, when a pH-active group is protonated the standard redox potential ($E^o$) of an adjacent redox-active group rises. Further, our results also show that E-REMD is able to achieve faster statistical convergence than CEMD or C(pH,E)MD. Moreover, computational benchmarks using our methodologies show high-performance of GPU accelerated calculations in comparison to conventional CPU calculations.

## I. INTRODUCTION

The coupling between electron and proton transfer is essential to describe many important biological processes. The protonation/redox state of proteins and other biomolecules is related to their structure and function, and it can affect properties like stability, ligand binding, catalysis, absorption spectrum, among others [1,2]. This happens because the solution's pH and redox potential (reduction potential or electrode potential) affect the charge distribution on the biomolecules due to changes in the predominant protonation/redox state of the titratable groups. For many systems, the standard redox potential (also called midpoint reduction potential) turns out to be pH-dependent [3,4]. Hence, theoretical methods that can correctly describe systems at constant pH and constant redox potential are very important. Previous works in this area have been published [5–12].

Due to the mathematical similarity between the Henderson-Hasselbalch equation (applied to acid-base reactions) and the Nernst equation (applied to electrochemistry, redox reactions), the theoretical derivations used on constant pH methods [13–16] can be extended to constant redox potential methods [17]. There are two types of Constant pH Molecular Dynamics (CpHMD) approaches [18]: the ones that consider continuous protonation states [19–22] and the ones that consider discrete protonation states [13,14]. The continuous protonation states approach is based on the introduction of a fictitious particle at each titratable site that is propagated through conventional Molecular Dynamics (MD) according to a pH-dependent force. The discrete protonation states approach makes use of Metropolis Monte Carlo exchange attempts between different protonation states over the course of the MD. Regardless of the approach type, the accuracy of predicted p$K_a$ values relies not only on the accuracy of the force field parameters but also on the extent of the conformational sampling.

Replica Exchange Molecular Dynamics (REMD) is a state-of-the-art method that is able to significantly improve conformational sampling convergence [23–28] while taking advantage of computational parallelization. This approach consists of individual simulations (which are considered as independent replicas) that periodically attempt to exchange information between them through a Metropolis Monte Carlo scheme. Studies regarding REMD along the pH dimension (pH-REMD) have been reported in the literature [18,21,24,25,29,30]. In pH-REMD, each replica explores the conformational space at a different pH value.

In this work, we present the implementation of constant redox potential MD (CEMD), of coupled constant pH and constant redox potential MD (C(pH,E)MD), and REMD along the redox potential dimension (E-REMD) in the AMBER software package [31]. To the best of our knowledge, this is the first implementation of E-REMD. Discrete redox and protonation states are considered in our methodologies, and calculations can be done using both implicit (i.e. Generalized Born, GB) or explicit solvent models. These new implementations are based on existing CpHMD and pH-REMD methods implemented on AMBER by Mongan, Swails, and others [13,18,25]. We present results for N-acetylmicroperoxidase-8 (NAcMP8) [32] with a histidine peptide as the axial ligand. NAcMP8 is a small peptide derived from cytochrome *c* that has a single ferric heme group and is a good reference compound candidate for the simulation of larger proteins containing one or more equivalent heme groups.

Previous simulations for large systems using AMBER's high-performance GPU code have shown to have great speedups over calculations using conventional CPUs [33,34]. Our CEMD, C(pH,E)MD and E-REMD implementations are also available using AMBER's GPU code. To our knowledge, these are the first implementations of constant redox potential methods using GPU accelerated code.

## II. THEORY AND METHODS

### A. Constant Redox Potential Molecular Dynamics (CEMD)

In CEMD we make use of Monte Carlo transitions between discrete redox states, represented by different atomic charge distributions for reduced and oxidized states of a given redox-active residue. Further, in CEMD a predetermined number of MD steps are performed, the simulation is then halted, and a redox state change is attempted. In order to understand how the energetic cost of this attempt is computed, we start from the reduction reaction:

$$A_{oxi}^{n+} + ve^- \xrightleftharpoons{K_e} A_{red}^{(n-v)} \tag{1}$$

The equilibrium constant is given by $K_e = \frac{[A_{red}^{(n-v)}]}{[A_{oxi}^{n+}][e^-]^v}$ and is unitless. It is possible to devise expressions equivalent to $pH = -\log[H^+]$ and $pK_a = -\log K_a$ for the redox potential $E = -\frac{k_b T}{F}\ln[e^-]$ and the standard redox potential $E^o = \frac{k_b T}{vF}\ln(K_e)$, where $F$ is the Faraday constant. Then, from the equilibrium constant expression we get to the Nernst equation [35]:

$$E = E^o + \frac{k_b T}{vF} \ln \frac{[A_{oxi}^{n+}]}{[A_{red}^{(n-v)}]} \tag{2}$$

From this equation, an expression for the fraction of reduced species is devised:

$$f_{red} = \frac{[A_{red}^{(n-v)}]}{[A_{red}^{(n-v)}] + [A_{oxi}^{n+}]} = \frac{1}{1 + e^{n\frac{vF}{k_b T}(E-E^o)}} \tag{3}$$

In this equation, $n$ is the Hill coefficient and is a post hoc correction added to account for the fact that the redox-active group might be affected by other pH- or redox-active groups nearby. We can also devise an expression for $f_{red}$ based on a statistical mechanics point of view:

$$f_{red} = \frac{\left[A_{red}^{(n-v)}\right]}{\left[A_{red}^{(n-v)}\right] + [A_{oxi}^{n+}]} = \frac{e^{\frac{-G_{reduced}}{k_bT}}}{e^{\frac{-G_{reduced}}{k_bT}} + e^{\frac{-G_{oxidized}}{k_bT}}} = \frac{1}{1 + e^{\frac{\Delta G_{reduction}}{k_bT}}} \quad (4)$$

The Gibbs Free Energy of reduction can be obtained from equations 3 and 4:

$$\Delta G_{reduction} = vF(E - E^o) \quad (5)$$

Any theoretical modeling based solely on this equation requires prior knowledge of the $E^o$ value, however this is one of the properties one wants to be able to predict theoretically. To overcome this issue, we also write equation 5 for a reference compound where the value of $E^o$ is known. We also split the Gibbs Free Energy of reduction into electrostatic (Coulombic) and non-electrostatic contributions:

$$\Delta G_{reduction} = vF(E - E^o) = \Delta G_{elec} + \Delta G_{non-elec} \quad (6)$$

$$\Delta G_{reduction,ref} = vF(E - E^o_{ref}) = \Delta G_{elec,ref} + \Delta G_{non-elec,ref} \quad (7)$$

The approximation made on AMBER is that $\Delta G_{non-elec} = \Delta G_{non-elec,ref}$, which leads to:

$$\Delta G_{reduction} = vF(E - E^o_{ref}) + \Delta G_{elec} - \Delta G_{elec,ref} \quad (8)$$

This is the equation used on CEMD to perform Monte Carlo redox state change attempts. $\Delta G_{elec,ref}$ is a precomputed quantity adjusted to reproduce the $E^o_{ref}$ value for the reference compound, and $\Delta G_{elec}$ is computed on every redox state change attempt. $\Delta G_{elec}$ is obtained by taking the difference between the electrostatic energy associated with the proposed and current redox states.

**B. Constant pH and Redox Potential Molecular Dynamics (C(pH,E)MD)**

In CpHMD, the core equation for protonation state change attempts [13,25,30], equivalent to equation 8, is:

$$\Delta G_{protonation} = k_bT(\ln 10)(pH - pK_{a,ref}) + \Delta G_{elec} - \Delta G_{elec,ref} \quad (9)$$

In our C(pH,E)MD implementation, protonation and redox state change attempts are performed separately, even if the attempts happen at the same MD step. The relation between protonation and redox processes comes naturally from the fact that a successful redox state change attempt will change the charges of a redox-active group and this change will thus affect the next protonation state change attempts of neighboring pH-active groups. Similarly, a successful protonation state change attempt will affect the next redox state change attempts of near redox-active groups.

## C. Replica Exchange Molecular Dynamics along the Redox Potential dimension (E-REMD)

The chemical potential $\mu_A$ associated to a substance $A$ is defined as $\mu_A = \mu_A^o + k_bT \ln a_A$, where $\mu_A^o$ is the standard chemical potential and $a_A$ is the activity of $A$. The standard chemical potential for electrons is chosen as $\mu_{e^-}^o = 0$, thus by making use of $E = -\frac{k_bT}{F}\ln[e^-]$ we obtain:

$$\mu_{e^-} = k_bT \ln[e^-] = -FE \quad (10)$$

This shows that performing simulations at constant redox potential is the same as performing simulations at constant chemical potential for electrons ($\mu_{e^-}$).

Hereafter, we show how the Monte Carlo exchange criterium derived by Itoh et al. [30] for pH-REMD can be extended to E-REMD. First, we consider a system of $M$ non-interacting replicas. Replica $i$ ($i = 1, ..., M$) has coordinate and momentum vectors $\vec{q}_i$ and $\vec{p}_i$, temperature T, a redox potential value of $E_l$ ($l = 1, ..., M$), and $N_i^{e^-}$ electrons. If we exchange two replicas, the following detailed balance condition must be satisfied:

$$P(X_i^l)P(X_j^n)w(X_j^n \to X_i^n, X_i^l \to X_j^l) =$$
$$P(X_i^n)P(X_j^l)w(X_j^l \to X_i^l, X_i^n \to X_j^n) \quad (11)$$

where $X_i^l \equiv (\vec{q}_i, \vec{p}_i, N_i^{e^-}, E_l)$, $P$ is the equilibrium probability, and $w$ is the transition probability. In the semi-grand canonical ensemble of electrons, where the volume $(V)$, temperature $(T)$, and $\mu_{e^-}$ are fixed, the equilibrium probability is given by:

$$P(X_i^l) = \frac{1}{\Xi} e^{[-\varepsilon_i + (\mu_{e^-})N_i^{e^-}]/k_bT} = \frac{1}{\Xi} e^{[-\varepsilon_i - FE_l N_i^{e^-}]/k_bT} \quad (12)$$

where $\varepsilon_i$ is the energy of the system, and $\Xi$ is the grand canonical partition function. We can write the ratio of the transition probabilities as:

$$\frac{w(X_j^n \to X_i^n, X_i^l \to X_j^l)}{w(X_j^l \to X_i^l, X_i^n \to X_j^n)} = e^{-\Delta} \quad (13)$$

$$\Delta = \frac{F}{k_bT}(E_n - E_l)(N_i^{e^-} - N_j^{e^-}) \quad (14)$$

The Metropolis Monte Carlo solution for the exchange probability can be then written as:

$$w(X_j^{m,n} \to X_i^{m,n}, X_i^{k,l} \to X_j^{k,l}) = \begin{cases} 1, & \Delta < 0 \\ e^{-\Delta}, & \Delta \geq 0 \end{cases} \quad (15)$$

It is important to notice that the acceptance ratio decreases exponentially with the difference in redox potential values. Therefore, it is important to choose redox potential values for the replicas such that the acceptance ratio is not too low. For example, for $T = 300\ K$, $N_i^{e^-} - N_j^{e^-} = 1$ and $E_n - E_l = 30$ mV we have $e^{-\Delta} \cong 31\%$. Roughly the same probability is obtained on pH-REMD by using pH intervals of 0.5. This is already expected because it can be shown that 1.0 pH unit is equivalent to 59.5 mV in redox potential units at $T = 300\ K$.

## D. Theoretical description of the pH dependence of $E^o$ and the redox potential dependence of $pK_a$ values

By making use of a thermodynamic cycle, it is possible to devise equations to describe the pH dependence of standard Redox Potential $E^o$ and the redox potential dependence of $pK_a$ values. This thermodynamic cycle and the derivations for all the equations shown in this subsection are available at the Supporting Information. By making thermodynamic considerations and assuming the system to contain only a single redox-active residue and one or more pH-active residues, we obtain:

$$vF\left(E^o_{prot} - E^o_{deprot}\right) = k_b T \ln(10) \sum_i \left(pK^{(i)}_{a,red} - pK^{(i)}_{a,oxi}\right) \quad (16)$$

where $E^o_{prot}$ and $E^o_{deprot}$ are the standard redox potential of the redox-active residue when all the pH-active residues are, respectively, fully protonated or fully deprotonated, and $pK^{(i)}_{a,red}$ and $pK^{(i)}_{a,oxi}$ are the $pK_a$ values of the $i^{th}$ pH-active residue when the redox-active residue is, respectively, fully reduced or fully oxidized.

Equation 16 shows that the difference between $E^o_{prot}$ and $E^o_{deprot}$ is directly related to differences between $pK^{(i)}_{a,red}$ and $pK^{(i)}_{a,oxi}$. As an example of that, if the pH-active groups don't suffer from the interaction with the redox-active group such that $pK^{(i)}_{a,red} = pK^{(i)}_{a,oxi}$ then according to equation 16 we must necessarily have $E^o_{prot} = E^o_{deprot}$.

We can also, in the following equation, describe the pH-dependent $E^o$ value for the case where the species of all pH-active residues are not all fully protonated nor all fully deprotonated:

$$E^o = E^o_{prot} + \frac{k_b T}{vF} \sum_i \ln\left(\frac{10^{-pK^{(i)}_{a,red}} + 10^{-pH}}{10^{-pK^{(i)}_{a,oxi}} + 10^{-pH}}\right) \qquad (17)$$

We observe that at the low pH limit (i.e., large $[H^+]$ values) $E^o$ becomes $E^o_{prot}$, and at the high pH limit (i.e., small $[H^+]$ values) $E^o$ becomes exactly the expression for $E^o_{deprot}$ that we obtain from equation 16.

The redox potential dependence of the $pK_a$ values of the pH-active residues can be described by the following equation:

$$\sum_i pK_a^{(i)} = \sum_i pK_{a,red}^{(i)} + \log\left(\frac{e^{-vFE^o_{prot}/k_bT} + e^{-FE/k_bT}}{e^{-vFE^o_{deprot}/k_bT} + e^{-FE/k_bT}}\right) \qquad (18)$$

where $pK_a^{(i)}$ is the redox potential dependent $pK_a$ value of the $i^{th}$ pH-active residue. In the limit of high redox potential values (i.e., small $[e^-]$ values) $pK_a^{(i)}$ becomes $pK_{a,oxi}^{(i)}$ and equation 18 turns into equation 16.

### III. CALCULATION DETAILS

In this work, all the simulations were performed using an in-house modified version of AMBER 16 [31]. Our implementations are part of the new AMBER 18 release. In this section we are going to discuss the C(pH,E)MD and E-REMD implementations, followed by details about parametrization of our test system, and the implicit and explicit solvent calculations performed.

### A. C(pH,E)MD implementation

Previous discrete protonation states CpHMD publications have analyzed, among other things, the behavior of $pK_a$ predictions in long Generalized Born (GB) implicit solvent simulations [18]. Also, how a higher frequency of exchange attempts affects convergence on pH-REMD [25], and how variables like the solvent relaxation time affects explicit solvent simulations [18]. These discussions will be revisited here in the context of constant redox potential, even though we expect to see these same behaviors due to the natural likeness between CpHMD and CEMD. This likeness can be seen by comparing equations 8 and 9 and arises from the similarities between the Henderson-Hasselbalch and the Nernst equations.

Figure 1 shows the workflow of C(pH,E)MD as implemented in AMBER. The number of standard MD steps between state change attempts is tunable and may be different for protonation and redox state change attempts. Thus, protonation and redox state change attempts may happen at the same MD step. In this case the protonation state change attempts are performed first.

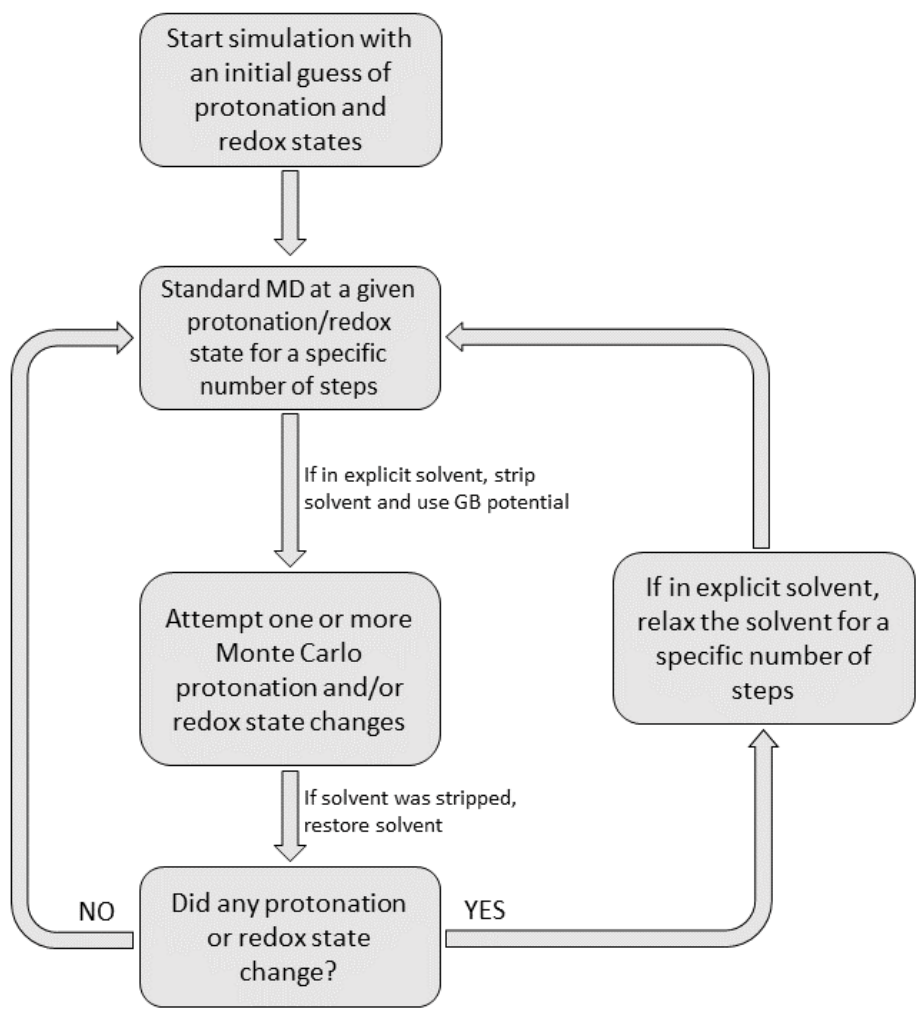

**Figure 1.** C(pH,E)MD workflow as implemented on AMBER. This workflow generalizes both the implicit and explicit solvent implementations.

In both the implicit and explicit solvent implementations, the proposed state of a residue in a protonation or redox state change attempt is chosen randomly from the available states of the residue, excluding the current state the residue is in.

**1. Implicit solvent (IS) implementation**

When the standard MD is halted for a protonation and/or redox state change attempt (see Figure 1), a single residue is picked randomly and is the only residue considered for a protonation or redox state change attempt.

## 2. Explicit solvent (ES) implementation

In the ES implementation, when the MD is halted one state change is attempted for each pH- and/or redox-active residue. The residues are chosen in random order until all pH- and/or redox-active residues are visited once. Likewise the ES CpHMD AMBER implementation [18], during redox state change attempts the water molecules and any eventual ions are stripped and the $\Delta G_{elec}$ term for a redox-active residue (see equation 8) is computed using a GB model. After the attempts for each residue are performed, the solvent molecules are restored. If any state change was accepted, then solvent relaxation is performed. Solvent relaxation consists on performing a predetermined number of MD steps only on the degrees of freedom associated to the solvent, including any possible ions, while the solute degrees of freedom are kept fixed. Solvent relaxation is done to adapt the solvent conformation to the new protonation and/or redox state. The number of solvent relaxation steps may be different for protonation and redox state change attempts. In order to make C(pH,E)MD more efficient, if a protonation and a redox state change attempt happen at the same MD step when the standard MD is halted (see Figure 1), only a single solvent relaxation is performed using either the number of relaxation steps for protonation or redox state change attempts, whichever is larger. This actually makes the computational cost of C(pH,E)MD close to CpHMD or CEMD in ES simulation, as shown in the section IV.D.

Doing one attempt for each residue when the simulation is halted allows us to perform protonation and redox state change attempts less frequently than in the IS implementation, thus lowering the total number of solvent relaxation steps over the course of the simulation. In ES constant pH and/or constant redox potential MD simulations, the solvent relaxation contributes the most to the computational cost of the methodology in comparison to regular MD. Therefore, lowering the total number of relaxation steps makes the simulation computationally more efficient.

During protonation or redox state change attempts, the sudden change in charge, even if the pH- or redox-active residue is not solvent exposed, destabilize the solvent molecules around the solute. This makes the use of ES calculations for state change attempts unfeasible as an unfavorable energy change would lead to a low state change probability. For this reason, our state change attempts require the use of implicit solvent calculations where the solvent instantaneously adapts to the new charge set. Stern [15] has proposed an ES CpHMD method that uses an interpolation between the current and the proposed protonation states and doesn't require the use of implicit solvent calculations, but its implementation has not been attempted in the present publication.

Another point worth to discuss is the fact that the net charge of the system changes when protonation or redox state change attempts are accepted. It has been shown that when periodic boundary conditions are in place and the electrostatic interactions are computed using a lattice sum, finite-size effects arise on simulations where the volume or the net charge is changing throughout the simulation [36,37]. These effects have been shown to be higher for small unit cells and could lead to unrealistic behaviors [36]. By using GB calculations for state change attempts in our methodology, we avoid these finite-size effects. In our implementation, the standard MD runs between state change attempts are done at constant charge.

**B. E-REMD implementation**

In replica exchange simulations, an important variable is the exchange attempt frequency (EAF). If the EAF equals zero, this means that no exchange attempts between replicas are performed and E-REMD becomes equivalent to C(pH,E)MD, or to CEMD if the constant pH option is not considered. Previous AMBER replica exchange publications have shown that by increasing the EAF the sampling convergence could be improved as the simulated system may overcome barriers more easily [25,38,39]. In this work, we show that a better convergence is also obtained for E-REMD on properties that depend on the redox potential in comparison to C(pH,E)MD or CEMD.

As can be seen in equations 14 and 15, the E-REMD exchange probability depends on the difference between redox potential values of the replicas we are trying to exchange. Therefore, in the same way that is done in AMBER for pH-REMD, in our E-REMD implementation we only attempt exchanges between replica neighbors to increase the exchange probability. To exemplify, consider 4 replicas ordered by increasing values of redox potential. At the first exchange attempt, we attempt to exchange replicas 1 with 2 and 3 with 4. On the next exchange attempt, the attempts are made between replicas 2 with 3 and 1 with 4. This cycle is repeated until the end of the simulation is reached.

## C. NAcMP8 parametrization

In many proteins containing heme groups, each iron atom at the center of the porphyrin rings is axially connected to two histidines [11,40,41]. For the theoretical modeling of such proteins, it is ideal to have a reference compound in agreement with this conformation. The standard redox potential of N-acetylmicroperoxidase-8 (NAcMP8) axially connected to an imidazole molecule has been experimentally measured as -203 mV *vs*. NHE at pH 7.0 [32]. NAcMP8 has a histidine residue from its peptide chain that is axially connected to the heme group. On the other side of the porphyrin plane, we axially connected a histidine peptide as shown in Figure 2. This closely represents the experimental situation.

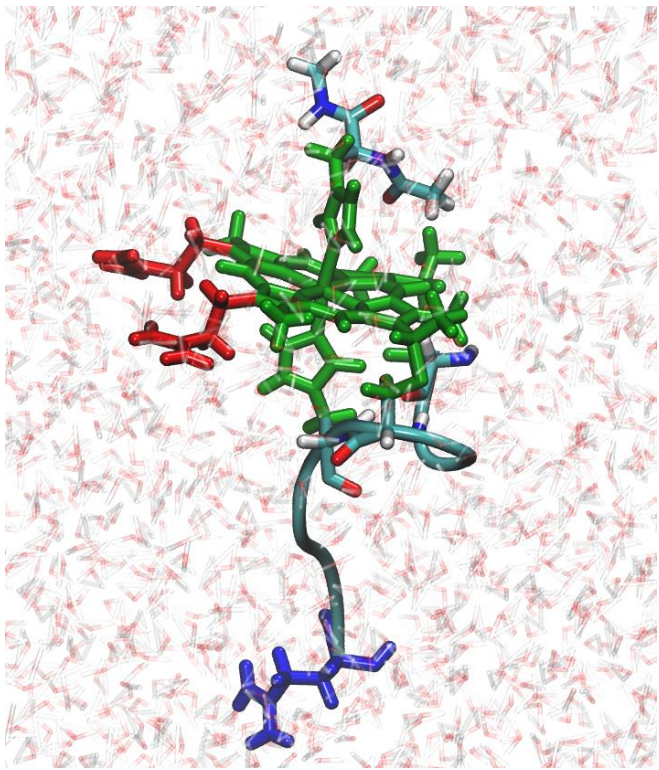

**Figure 2.** N-acetylmicroperoxidase-8 (NAcMP8) with a histidine peptide as the axial ligand. The residue HEH is shown in green and the two propionates are shown in red. The glutamate is shown in blue.

The structure of NAcMP8 is sketched in the Figure 1 of reference [32]. From this figure and by making modifications to the structure of the horse heart cytochrome *c* [42] available at the PDB code 1HRC, we obtained the structure shown in Figure 2. The iron atom, the porphyrin ring, together with the side chains of two histidines and two cysteines were considered as a single residue we called HEH, which is colored in green at Figure 2. HEH is the redox-active residue that changes its atomic charge distribution when a redox state change attempt is successful. Therefore, a redox state change affects the charge distribution on the histidines and cysteines.

The charge distributions of the reduced and oxidized states of HEH can be found at Table S1 in the Supporting Information. NAcMP8 contains three pH-active residues: two propionates (PRN) colored in red in Figure 2, and one glutamate (GL4) colored in blue. The propionates are separate residues from HEH. The charge distributions of the deprotonated and protonated states of PRN can be found at Table S2.

Residues other than the heme group were parametrized using the AMBER FF99SB force field [43]. The heme force field parameters were taken from Crespo *et al.* [44], with the atomic charges for the reduced and oxidized states adapted from Henriques *et al.* [40]. The experimental p$K_a$ of a propionic acid is 4.85 [45]. The $\Delta G_{elec,ref}$ term from equation 9 was fitted in implicit and explicit solvent constant pH simulations of a propionic acid in water using $pK_{a,ref} = 4.85$. The fitted value was then used in the simulations for the NAcMP8 propionates. $\Delta G_{elec,ref}$ from equation 8 for the HEH residue was also fitted in IS and ES using constant pH and redox potential simulations of NAcMP8 with a histidine peptide as the axial ligand with pH = 7.0 and $E^o_{ref} = -203$ mV. The glutamate residue has $pK_{a,ref} = 4.4$ and we used values already available in AMBER's *cpinutil.py* tool for its $\Delta G_{elec,ref}$ in IS and ES [25].

In the explicit solvent calculations, the intrinsic solvent radius of the carboxylate oxygens on the propionates and the glutamate was changed from 1.5 to 1.3 Å in order to compensate for having two dummy hydrogens present on each oxygen [18,46].

**D. Implicit Solvent Calculations**

We begin with the initial structure having the heme group in the oxidized state, and both propionates and the glutamate in the deprotonated state. This structure is then minimized for 100 steps using the steepest descent algorithm and then for 3900 steps using the conjugate gradient algorithm constraining the backbone atoms with a 10 kcal/mol·Å² constant. The minimized structure is then heated during 3 ns by varying linearly the target temperature from 10 to 300 K over the initial 0.6 ns. During heating, the backbone atoms were constrained using a constant of 1 kcal/mol·Å², and the temperature was controlled using Langevin dynamics with a friction frequency of 5 ps$^{-1}$.

An equilibration at 300 K is then performed on the heated structure for 10 ns using Langevin dynamics with a 10 ps$^{-1}$ friction frequency and a 0.1 kcal/mol·Å² constrain on the backbone atoms. This equilibrated structure was used as the initial structure for the production simulations from where our results were extracted. For the production simulations, no positional restraints are applied and redox and/or protonation state change attempts are performed every 10 fs. Calculations with E-REMD were done using an exchange attempt frequency (EAF) of either 0.5 or 50 ps$^{-1}$, meaning one exchange attempt every 2000 or 20 fs respectively. All simulations were done with a time step of 2 fs and all bond lengths of bonds containing hydrogen atoms were constrained using the SHAKE algorithm [47,48].

Previous AMBER implicit solvent CpHMD publications [13,18,25] have used the Generalized Born model proposed by Onufriev *et al.* [49] (represented by the input flag *igb*=2 on AMBER) to account for solvent effects. For consistency, the same model was used in our implicit solvent simulations, both during MD and during protonation and redox state change attempts.

**E. Explicit Solvent Calculations**

In the initial structure, the heme group is in the oxidized state, and both propionates and the glutamate are in the deprotonated state. The system was solvated using TIP3P waters in a truncated octahedron box with a buffer of 10 Å (distance between the wall and the closest atom in the solute). In order to neutralize the total charge at this initial state, two sodium ions were randomly added to the solution.

The initial structure was minimized, constraining the backbone atoms with a 10 kcal/mol·Å$^2$ constant, for 1000 steps using the steepest descent algorithm followed by 4000 steps of conjugate gradient. The minimized structure was then heated at constant volume during 3 ns using a backbone constrain of 1 kcal/mol·Å$^2$ and Langevin dynamics with a friction frequency of 5 ps$^{-1}$ to control the temperature. During heating, the target temperature was linearly varied from 10 to 300 K over the initial 0.67 ns.

The heated structure was then equilibrated at constant pressure (1 bar) and temperature (300 K) for 8 ns with no backbone restraints, using a friction coefficient of 2 ps$^{-1}$ for the Langevin dynamics to maintain the temperature and relaxation time of 3 ps for the Berendsen barostat to control the pressure. The system was then submitted to a new equilibration at constant volume and temperature (300 K) for 20 ns using Langevin dynamics with a 2 ps$^{-1}$ friction frequency. The structure from this equilibration was used as input for the production simulations from where our results were extracted.

It has been shown for ES CpHMD that when a protonation state change is accepted, relaxation times up to 4 ps could be required to completely stabilize the solvent, however, the most significant part of the solvent stabilization is done during the first 200 fs [18]. As relaxation times on the order of 4 ps would have very significant effects on the computational cost of the methodology, lower computational relaxation times of around 200 fs were attempted. No significant differences were observed in the p$K_a$ predictions in comparison to higher relaxation times of up to 2 ps. We verified this same behavior in ES CEMD or C(pH,E)MD on E° predictions.

In our production simulations, redox and/or protonation state change attempts are performed every 200 fs, and the solvent relaxation is performed for 200 fs. The E-REMD calculations were performed using an Exchange Attempt Frequency (EAF) of either 0.5 or 5 ps$^{-1}$; this means one exchange attempt every 2000 or 200 fs respectively. The time step used on all MD simulations, including during solvent relaxation, is 2 fs and all bonds containing hydrogen atoms were constrained using the SHAKE algorithm [47,48]. The particle-mesh Ewald method [50,51] using a van der Waals cutoff and a direct space of 8 Å was considered for the long-range electrostatic interactions.

As discussed in the section III.A, the redox and protonation state change attempts are performed in implicit solvent. In our explicit solvent calculations, we also use the GB model proposed by Onufriev *et al.* [49] (*igb*=2 on AMBER) for that.

**IV. RESULTS AND DISCUSSIONS**

As discussed before, replica exchange simulations are meant to accelerate convergence. This means a REMD simulation should converge faster than a simulation without REMD for the same number of steps. We will now test this concept for E-REMD by presenting results for short simulations in which neither the CEMD nor C(pH,E)MD results are converged, and also evaluate what happens as a function of simulation time on long simulations. We then analyze how the use of E-REMD affects the results.

## A. CEMD vs E-REMD

### 1. Short simulations

CEMD results are presented for simulations in which the propionates and the glutamate are both either protonated or deprotonated throughout the whole simulation, including during minimization, heating and equilibration. Production simulations were executed for only 50 ps (25,000 MD steps) in IS and 1000 ps (500,000 MD steps) in ES. As discussed previously, we are making protonation and redox state change attempts every 10 fs in IS and every 200 fs in ES. For this reason, significantly more state change attempts are performed in IS than in ES during the same simulated time interval. In order to account for that, we are performing more steps for ES than for IS in this analysis.

Production simulations were performed for 12 values of redox potential, using CEMD and also E-REMD with two different EAFs. Figure 3 shows the fraction of reduced species of HEH as a function of E, and the fittings of $E^o$ and Hill coefficient using equation 3 to the data obtained from the simulations. The errors reported for $E^o$ and Hill coefficient are the errors in the fitting to equation 3. Results are shown for both implicit and explicit solvent models.

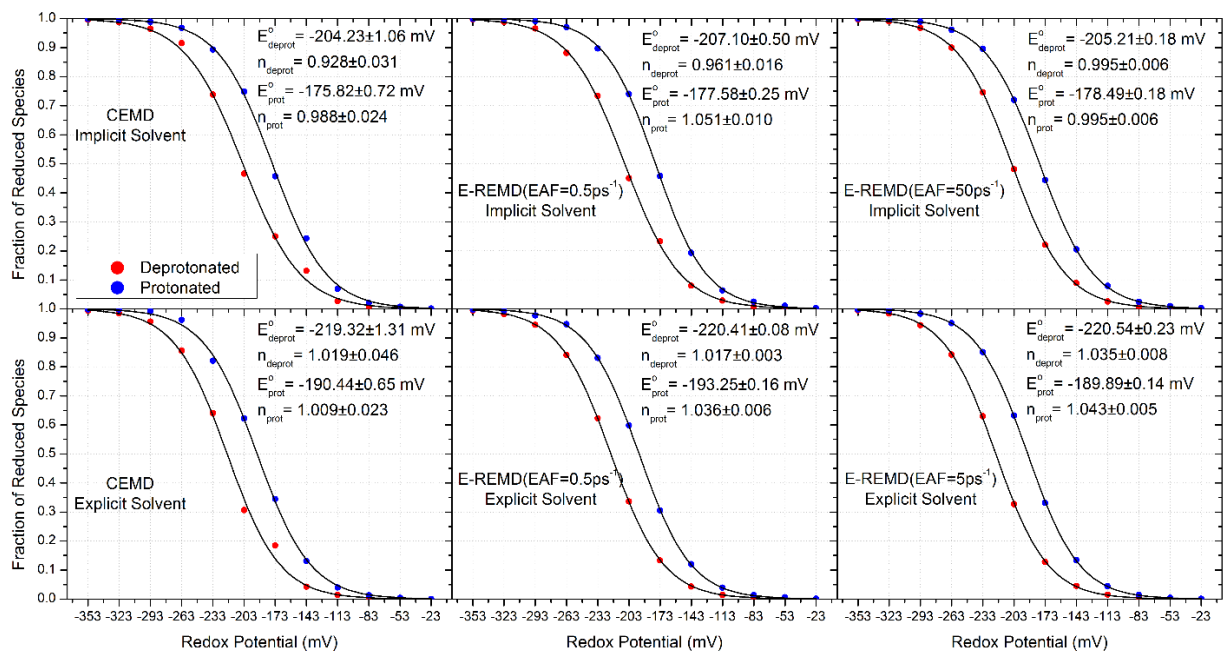

**Figure 3.** Fraction of reduced species of HEH as a function of the redox potential for short constant redox potential simulations: 50 ps in IS and 1000 ps in ES. The pH-active residues are either all deprotonated or protonated throughout the whole simulation. EAF = Exchange Attempt Frequency.

The fractions of reduced species obtained from the simulations are in agreement with the expected behavior predicted by equation 3. Lower fitting errors are obtained when replica exchange is used.

A protonation of a pH-active group leads to a positive increase on its charge. Hence, as electrons can neutralize the positive charge of protons, the standard redox potential of an adjacent redox-active group increases as it becomes easier for an electron to be attached there. The standard redox potentials shown in Figure 3 have more positive $E^o$ values when all pH-active residues are protonated in comparison to when they are all deprotonated, therefore in agreement with the expectation.

## B. C(pH,E)MD vs E-REMD

### 1. Short simulations

We now extend the same test from the previous section to C(pH,E)MD, and, starting from the equilibrated structure where the propionates and the glutamate are deprotonated, we let their protonation states change during the production simulations while still allowing the redox states to change as before. Here, production simulations were executed for only 50 ps in IS and 1000 ps in ES, and for 18 different pH values ranging from 2.0 to 10.5 in intervals of 0.5. Further, for each pH value, 12 values of redox potential from $-353$ to $-23$ mV with interval of 30 mV were considered. Results are shown for C(pH,E)MD and E-REMD with two different EAFs, where in E-REMD only replicas with the same pH are allowed to have their redox potential values exchanged. For each pH, we plotted the fraction of reduced species versus redox potential (as in Figure 3) to extract the value of $E^o$. Then, the $E^o$ of HEH as a function of pH is shown in Figure 4. Results are shown for both implicit and explicit solvent models.

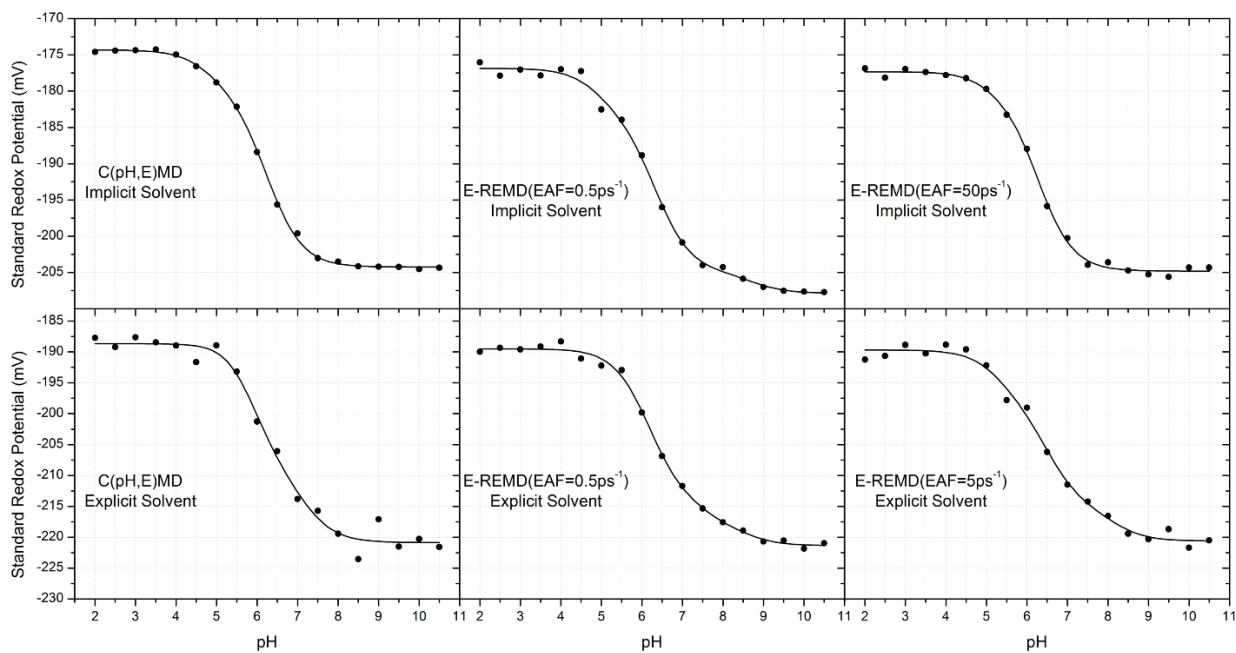

**Figure 4.** Standard redox potential (E°) of HEH as a function of pH for short simulations: 50 ps in IS and 1000 ps in ES. Dots are from the simulations and the solid curves are the fittings using equation 17. EAF = Exchange Attempt Frequency.

As the pH of the solution increases, the concentration of deprotonated species increases, leading to lower values of E° as it becomes more energetically unfavorable for an electron to reduce the heme group. This behavior has been reported experimentally [3,4]. Further, it can be seen that the most significant changes in E° with respect to pH happen for pH values around the p$K_a$s of the pH-active residues that more closely interact with the redox-active residue. The results shown in Figure 4 are in agreement with these trends. The solid curves are the fittings of the data from the simulations to equation 17. As can be seen in figure 4, this equation clearly describes the pH-dependence of E° values observed in our simulations. Also, the low and high pH limits E° values matches with the E° values predicted in the previous section using CEMD for all protonated and all deprotonated pH-active residues, respectively.

## 2. Long simulations

Here, we analyze how our standard redox potential predictions behave as a function of simulation time for C(pH,E)MD and E-REMD. In some situations, like for the study of processes that happen at long time scales, long simulations must be performed and, due to computational efficiency limitations, GPU-accelerated CUDA calculations must be used. Therefore, the analysis to be shown here demonstrate if our methodologies are stable and can be used in these situations.

For this analysis, production simulations were executed for 300 ns in both IS and ES. Both the Single Precision Fixed Point (SPFP) and the Double Precision Fixed Point (DPFP) AMBER 16 CUDA precision models [52] were used for the IS calculations, and only SPFP was used for the ES calculations. By construction, the CUDA DPFP implementation contains no further approximations in comparison to the CPU code and is the precision model used in AMBER to directly compare CPU and GPU results. Simulations were performed for pH = 7.0 and 12 redox potential values from -353 to -23 mV. The cumulative fraction of reduced species for each redox potential value was obtained as a function of time, and for each time we gathered all the fractions to fit $E^o$ using equation 3. The cumulative prediction of the $E^o$ of HEH as function of simulation time can then be obtained. This procedure was independently repeated 5 times. Figure 5 contains the average for the 5 independent simulations of the cumulative $E^o$ as a function of simulation time. We also report the standard deviation of the $E^o$ predictions in the 5 independent simulations as a function of time. Results are presented for C(pH,E)MD and E-REMD with three different EAFs, for both IS and ES calculations.

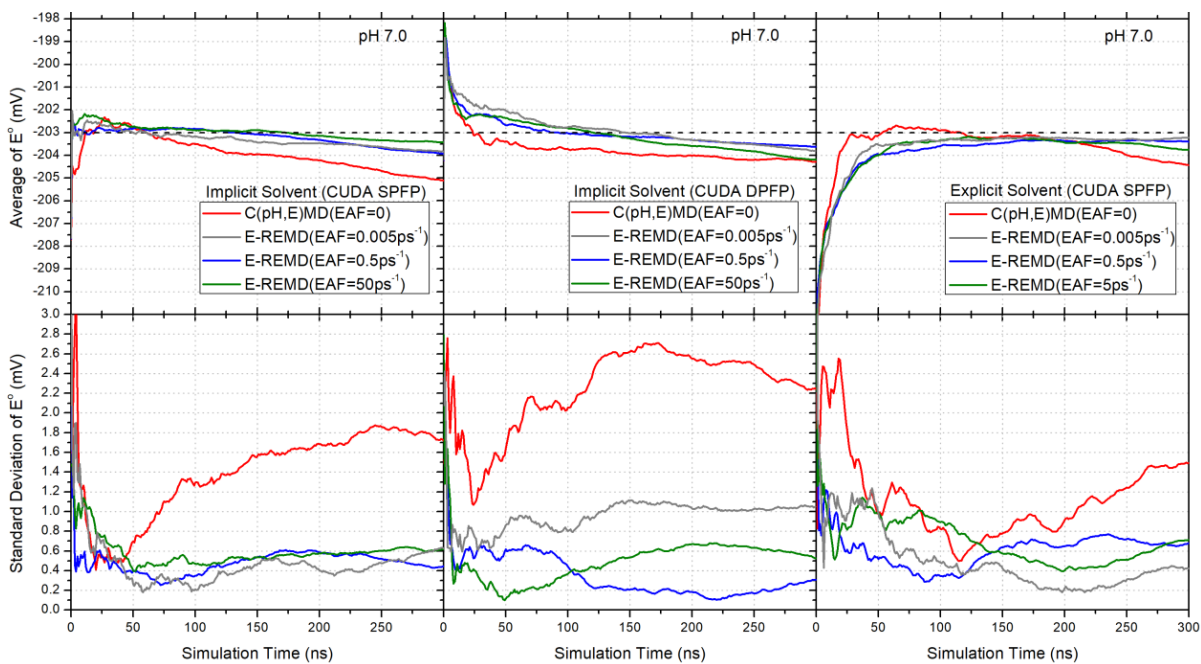

**Figure 5.** Cumulative standard redox potential (E°) of HEH averaged for 5 independent simulations (top) and its standard deviation (bottom) as a function of time for pH 7.0. EAF = Exchange Attempt Frequency; SPFP = Single Precision Fixed Point; DPFP = Double Precision Fixed Point.

The standard deviations in the E° predictions can be interpreted as an error measure of our methodologies. For both implicit and explicit solvent simulations, we observe that the C(pH,E)MD and E-REMD predictions agree with each other within the errors reported. The significantly lower errors for E-REMD in comparison to C(pH,E)MD are a very compelling evidence of the better convergence efficiency of E-REMD. It is also important to emphasize that the error bars obtained are reasonably small. We see error bars of up to 2.7 mV for C(pH,E)MD and of up to 1.2 mV for E-REMD. At 300 K these values correspond to 0.045 and 0.020 in pH units, respectively.

To complement the analyses done from Figure 5, in Figure 6 we split the simulations into 5 ns chunks and compute the predicted E° of each window using the fractions of reduced species for all redox potential values. Figure 6 was also generated using data from the 5 independent runs.

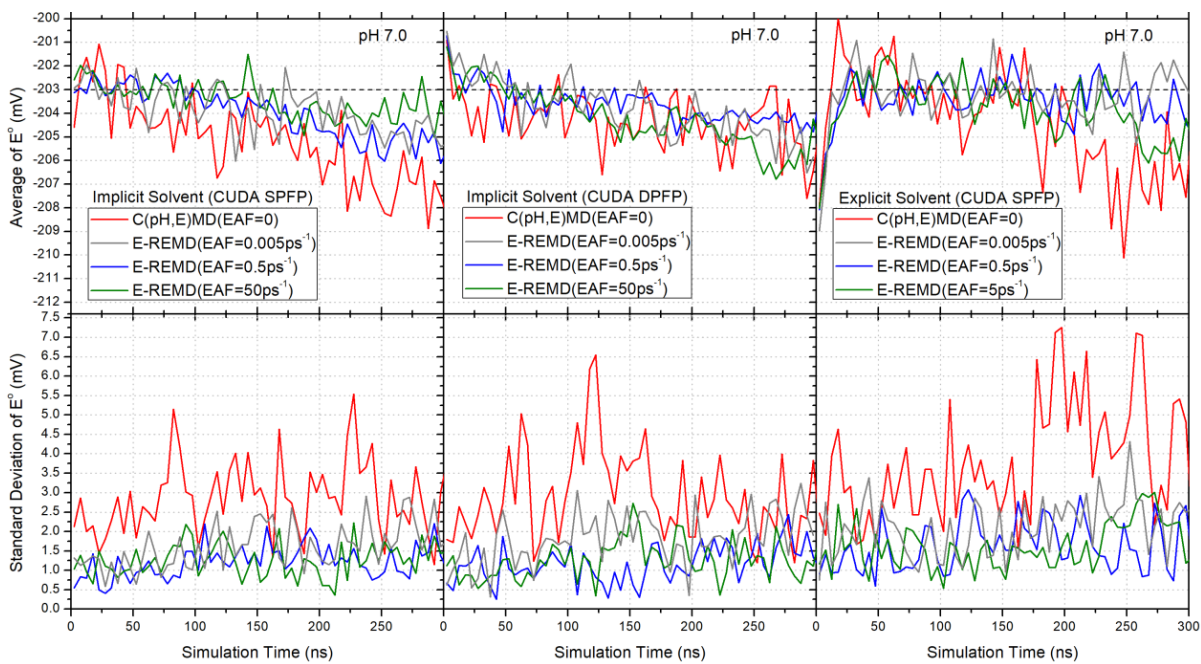

**Figure 6.** Chunk standard redox potential (E°) of HEH averaged for 5 independent simulations (top) and its standard deviation (bottom) computed for windows of 5 ns as a function of time for pH 7.0. EAF = Exchange Attempt Frequency; SPFP = Single Precision Fixed Point; DPFP = Double Precision Fixed Point.

Figure 6 also confirms smaller standard deviations for E-REMD simulations in comparison to C(pH,E)MD. In some cases, we also observe that the chunk E° predictions are clearly different at the beginning and at the end of the simulation, which means the chunk E° values do not fluctuate anymore around the target value of -203 mV. This explains the E° drops observed in Figure 5 generally after 200 ns of simulation. In order to understand this behavior, we looked at one of the C(pH,E)MD IS SPFP independent runs. For this simulation, we concluded the E° drop is caused by a conformational change (a flip) in the histidine that belongs to the NAcMP8 peptide chain.

The side chain of this histidine is part of the HEH residue (see Figure 2). More details and discussions about this are provided at section IV of the Supporting Information. As we show in the Supporting Information, even if this conformational change happens for only one of the redox potential values close to $E^o$, this is already enough to affect the $E^o$ fitting. In the simulation analyzed, this conformational change happened only once throughout the 300 ns simulated. Therefore, the addition of redox potential replica exchange helps to mix different conformations across different target redox potential values, which possibly explains the fact that the $E^o$ drops in Figures 5 and 6 are more subtle for E-REMD than for C(pH,E)MD.

Previous AMBER replica exchange publications have shown that increasing EAF in T-REMD simulations improves the convergence in structural analyzes due to the better sampling of the conformational space [39]. For pH-REMD, where a better sampling of the pH-space is done, better convergence in the p$K_a$ prediction for some residues are obtained when EAF is increased [25]. In this pH-REMD paper this conclusion about increasing EAF is reached mostly based in the results for a single residue, Asp 66. The authors, Swails *et al.*, attribute this behavior for Asp 66 to a better mobility of flexible regions of the protein they studied (HEWL). As can be seen in their paper, excluding Asp 48, all the other pH-active residues are effectively insensitive to increasing the EAF.

As can be seen in both Figures 5 and 6, the standard deviations in the E-REMD simulations do not change significantly when we increase EAF from 0.005 ps$^{-1}$ to higher values. C(pH,E)MD corresponds to the limit of EAF equal to zero, therefore if we decrease EAF to values below 0.005 ps$^{-1}$ we must at some point start seeing the same behavior observed for C(pH,E)MD. However, as can be seen, even the small EAF of 0.005 ps$^{-1}$ (where 1500 exchange attempts are performed during 300 ns of simulation) was enough to see effectively the same behavior observed for the higher EAF used.

Another important discussion to be done here has to do with the fact that systems containing at least a single heme group were not included during the parametrization of the GB model proposed by Onufriev *et al.* [49] (*igb*=2 on AMBER) used in this paper. Thus, the iron ion is the only heme atom without specific GB parameters. In our simulations we used the values 1.5 and 0.8 Å for the iron GB radius and screen, respectively. In the explicit solvent simulations, we see that the water molecules are always more than 4 Å apart from the iron atom, as can be seen at Figure S2 at the Supporting Information. As the iron ion is then nearly buried, one would expect that the exact values for these GB parameters will not influence the simulation results significantly. Also, it is important to check whether the porphyrin ring and the iron ion remain planar over the course of the implicit simulations. As the data in section II of the Supporting Information shows, we observe that both the porphyrin ring and the iron ion remain planar in both the implicit and the explicit solvent simulations. The dihedral angles analyzed obtained in implicit solvent are close to the values obtained in explicit solvent.

## C. Predicting Eº vs pH and p$K_a$ vs E

As discussed in the section IV.B.1, Eº should decrease with the increase of the solution pH. Following the same reasoning, equivalently the p$K_a$ of a pH-active residue should also decrease with the increase of the solution redox potential. Based on the discussions from Figures 5 and 6, we present 140 ns production simulation results in both IS with CUDA DPFP and in ES with CUDA SPFP. We have used E-REMD with EAF = 5 ps$^{-1}$ in ES and EAF = 50 ps$^{-1}$ in IS, 18 different pH values from 2.0 to 10.5, and for each pH value 12 values of redox potential from

−353 to −23 mV were considered. Figure 7 shows the standard redox potential of HEH as a function of pH, and the p$K_a$ of all pH-active residues and their sum as a function of the redox potential for both implicit and explicit solvent simulations. The solid lines in the figure are fittings using equations 17 and 18.

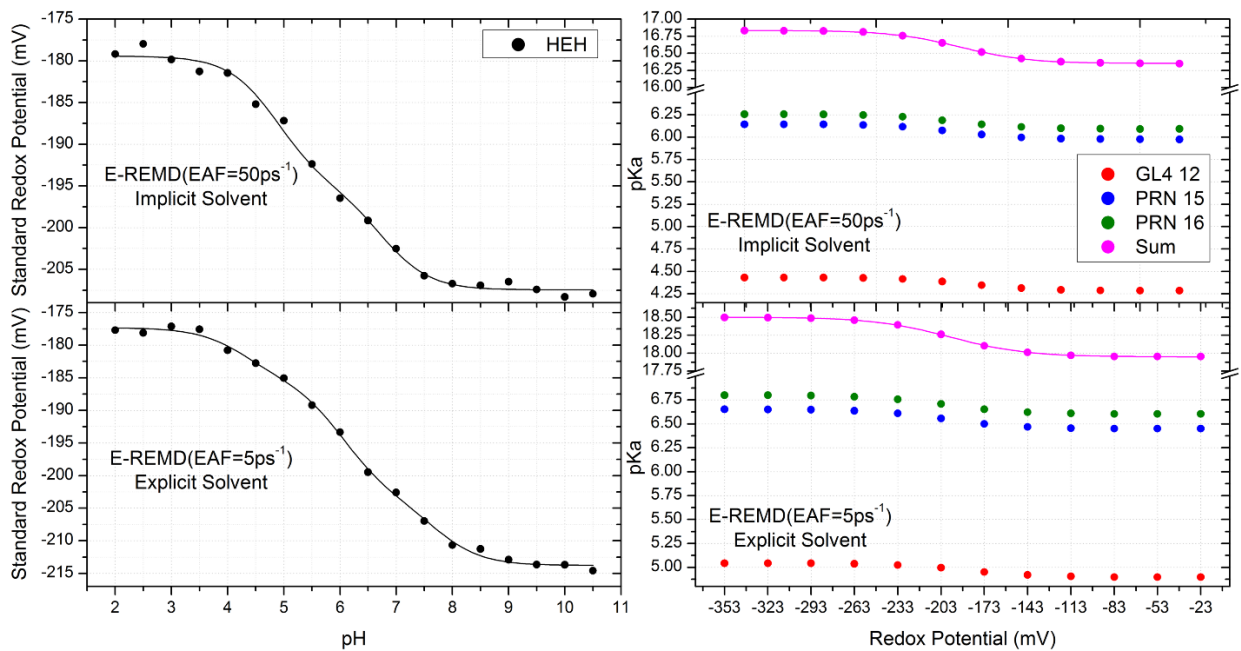

**Figure 7.** Standard redox potential (E°) of HEH as a function of pH, and p$K_a$ values of the pH-active residues as a function of the redox potential. The solid lines are fittings using equations 17 and 18 to the data from the simulations. E-REMD simulations are 140 ns long in IS (CUDA DPFP) and ES (CUDA SPFP) respectively. EAF = Exchange Attempt Frequency; GL4 = Glutamate; PRN = Propionate

As Figure 7 shows, we obtain a very good agreement between the theoretical predictions from equations 17 and 18 and the data from the simulations. It is important to mention that if we simply get the parameters in the equations ($E^o_{prot}$, $E^o_{deprot}$, $pK^{(i)}_{a,red}$ and $pK^{(i)}_{a,oxi}$) from the simulations, thus without fitting, we already obtain a good description of the simulation data shown in Figure 7. We also observe, as one would expect from the equations, that the pH values for which the E° of HEH changes the most are around the p$K_a$ values of the pH-active groups that more closely interact with the heme group. Equivalently, the p$K_a$ of each pH-active residue changes the most for redox potential values around the E° of HEH. As the figure shows, the p$K_a$ values of the two propionates are not exactly the same. They actually differ by ~ 0.15 pH units. The p$K_a$ values for the limits of low and high redox potential values differ by 0.2 pH unit for each propionate and by 0.15 pH unit for the glutamic acid. Therefore, even though the glutamic acid is farther apart from the heme than the propionates, it still significantly interacts with the heme group. Thus, for NAcMP8, the pH dependence of the standard redox potential of the heme group cannot be completely explained by considering just a single pH-active site or just the heme propionates. As an example, for the implicit solvent simulation, using equation 16 we see that the GL4 residue contributes with 8.8 mV to the difference of 28.6 mV between the E° values for the limits when the pH-active residues are fully protonated and fully deprotonated.

As both in the implicit and explicit solvent calculations the protonation and redox state change attempts were performed using the same GB model, the differences between our IS and ES simulations on the relative E° and p$K_a$ predictions in comparison to the reference compound can only come from the different ensemble of configurations generated in both cases. Figure 7 shows these conformational differences are significant enough to produce different E° and p$K_a$ values in IS and ES, even though the overall trends are the same. The p$K_a$ predictions are ~ 0.5 pH units higher in ES than in IS. Also, even though the E° values at low pH limits are basically the same, at the high pH limit E° is ~ 12 mV lower in ES than in IS. In addition, as one would expect from Figure 5, we can also observe the influence of the total number of MD steps by comparing Figure 7 with the results for short simulations shown in Figure 4.

For consistency, it is important to compare the low and high pH limiting E° values with the E° values predicted using E-REMD for the same EAF and same total number of steps but without the constant pH option. Without using constant pH, the E° values obtained when all pH-active residues are protonated and deprotonated are respectively -178.55 and -207.95 mV in IS, and -176.98 and -214.86 mV in ES. When constant pH is used, the E° values for the low and high pH limits are respectively -179.18 and -207.93 mV in IS and -177.69 and -214.58 mV in ES. The agreement between the E° predictions is good, as the predictions agree within less than 1 mV.

By considering a model where the pH-dependence of the heme group is associated to a single pH-active group, Das and Medhi [53] were able to experimentally infer for microperoxidase 11 the p$K_a$ values that correspond to the heme being fully reduced and fully oxidized. Microperoxidase 11 differs from NAcMP8 for having an extended peptide chain which contains a pH-active lysine reside. Also, the axial ligand in their experiments is a water molecule, instead of a histidine as is our case. Using their single pH-residue model for the range of pH values between 5.0 and 7.5 they obtained p$K_a$ values of 7.1 and 6.3 when the heme group is reduced and oxidized, respectively. According to equation 16, the difference between these p$K_a$ values should be directly related to the difference of E° values for when the pH-active group is fully protonated and fully deprotonated. From our ES simulations, the p$K_a$ values for both propionates at the low and high redox potential limits corresponding to the heme being fully reduced and fully oxidized are, respectively, 6.65 and 6.45 for PNR 15 and 6.80 and 6.60 for PNR 16. Assuming the experimental p$K_a$ values of 7.1 and 6.3 to correspond to the heme propionates, we see that the propionate p$K_a$s values obtained in our simulations are in the same range as the experimental ones.

In another publication [54], Das and Medhi were also able to experimentally infer for several different proteins the p$K_a$ values that would correspond to the heme propionates. For cytochrome $c_2$, they obtained p$K_a$ values of 7.4 and 6.3 when the heme group is reduced and oxidized, respectively. For cytochrome $b_5$, where the propionates are more solvent exposed, they obtained p$K_a$ values are 5.9 and 5.7. We see the propionate p$K_a$ values obtained in our simulations are in the same range as the experimental ones for cytochrome $c_2$. As can be seen in Figure 2, the propionates on NAcMP8 are solvent exposed. As Figure 7 shows, the Δp$K_a$ for each propionate when the heme group fully reduced and fully oxidized is 0.2. This difference matches with the one obtained experimentally for cytochrome $b_5$ where the propionates are mostly solvent exposed. However, this comparison should not be over interpreted as in the experimental paper only a single pH-active residue was considered in their model, where in our simulations 3 pH-active residues are considered.

**D. Computational Benchmarks**

In Table 1 the computational efficiency of the *pmemd* and *sander* AMBER modules is compared for C(pH,E)MD. Calculations were done using Cray XK7 nodes (that have Tesla K20X GPUs) at the Blue Waters supercomputer. In implicit solvent, there are 237 atoms (corresponding to the NAcMP8 and the histidine peptide) and in explicit solvent this number is 7403 atoms (there are 2388 water molecules and 2 Na$^+$ ions). The calculations were performed for pH = 7.0 and E = −203 mV.

|                          | **Implicit Solvent** | **Explicit Solvent** |
|--------------------------|:---:|:---:|
| **Computation**          | **Computational performance (ns/day)** ||
| *sander* Serial (1 CPU)   | 6.38   | 0.27   |
| *sander* MPI (2 CPUs)     | 13.18  | 0.55   |
| *sander* MPI (4 CPUs)     | 25.11  | 1.04   |
| *sander* MPI (8 CPUs)     | 42.98  | 1.71   |
| *sander* MPI (16 CPUs)    | 75.44  | 2.78   |
| *pmemd* Serial (1 CPU)    | 4.48   | 0.52   |
| *pmemd* MPI (2 CPUs)      | 9.40   | 1.02   |
| *pmemd* MPI (4 CPUs)      | 18.11  | 1.99   |
| *pmemd* MPI (8 CPUs)      | 31.42  | 3.30   |
| *pmemd* MPI (16 CPUs)     | 56.95  | 5.24   |
| *pmemd* CUDA SPFP (1 GPU) | 336.58 | 126.56 |
| *pmemd* CUDA DPFP (1 GPU) | 135.23 | 59.34  |

**Table 1.** Computational performances of C(pH,E)MD for different *sander* and *pmemd* computations.

*sander* was the first module implemented on AMBER capable of performing molecular dynamics. *pmemd* started as a reimplementation of some *sander* functionalities, intended to improve the computational performance of the simulations. In Table 1, we see that for serial and MPI the IS C(pH,E)MD calculations with *sander* are faster than with *pmemd*, however for ES, where the total number of atoms increases 31 times in comparison to the IS calculations, the computational performance is better for *pmemd* than for *sander*. This shows the scalability of *pmemd* is better, which means it would perform even better for larger systems in comparison to *sander*. Even though the calculations scale well with the number of CPUs, Table 1 also emphasizes the high-performance aspect of the GPU enabled implementation. Great speedups are observed in both IS and ES calculations. In ES, we see that the SPFP calculation using GPU is 243 times faster than the serial calculation and 24 times faster than the fastest MPI calculation. The SPFP precision model is around 2 times faster than the DPFP precision model in the CUDA calculations.

It is important to mention that the computational cost of our methodology in ES depends on the values of E and pH. If the redox potential value is close to the E° of a redox-active residue or if the pH value is close to the p$K_a$ of a pH-active residue, the probability of accepting a state change attempt increases, and in explicit solvent this means that more relaxation steps will be performed.

In Table 2 we compare the computational cost of E-REMD, C(pH,E)MD, CpHMD, and CEMD in comparison to regular MD. Here we have only used the *pmemd.cuda_SPFP.MPI* module. For C(pH,E)MD we performed calculations for 36 (pH,E) values, combining $pHs = 3.5, 4.0, 4.5, 5.0, 5.5$ and $6.0$ with $Es = -263, -233, -203, -173, -143$ and $-113$ mV. For CpHMD, CEMD and regular MD we have executed the same simulations as in C(pH,E)MD but turning off respectively the constant redox potential, the constant pH, and both the constant redox potential and pH. For E-REMD we have performed the same simulations as in C(pH,E)MD but allowing the redox potential values between replicas of same pH to be exchanged. The computational performances shown in Table 2 are the averages for the 36 (pH,E) simulations.

| Calculation | Implicit Solvent | Explicit Solvent |
|---|---|---|
|  | Computational performance (ns/day) | |
| Regular MD | 491.86 | 225.32 |
| CpHMD | 391.55 | 154.74 |
| CEMD | 389.62 | 163.83 |
| C(pH,E)MD | 289.13 | 145.08 |
| E-REMD at CpH (EAF = 0.5 ps$^{-1}$) | 275.68 | 103.54 |
| E-REMD at CpH (EAF = 50 ps$^{-1}$ IS and 5 ps$^{-1}$ ES) | 198.10 | 96.81 |

**Table 2.** Computational performances of *pmemd.cuda_SPFP.MPI* for different calculations.

As Table 2 shows, the addition of constant pH and/or constant redox potential has a bearable computational cost in comparison to regular MD. CpHMD and CEMD are 21% slower in IS and around 30% slower in ES than regular MD. In IS, the computational performances of CpHMD and CEMD are essentially the same. In ES, because each residue is visited once during protonation or redox state change attempts, the small difference between CpHMD and CEMD performances can be explained by the fact that we have three pH-active residues and only one redox-active residue. C(pH,E)MD is 41% slower in IS and 36% slower in ES than regular MD. We observe that the C(pH,E)MD performance is close to the CpHMD and CEMD performances for ES.

At the section IV.B we discussed the effect of increasing the EAF in our E-REMD simulations. As can be seen at Table 2, the computational cost of the simulation raises when EAF is increased, mainly in implicit solvent. The addition of replica exchange has a bearable computational cost in comparison to C(pH,E)MD. For the largest EAF we used, E-REMD is 31% slower in IS and 33% slower in ES than C(pH,E)MD.

## V. CONCLUSIONS

Based on the existing CpHMD and pH-REMD methods [13,18,25] implemented by Mongan, Swails, and others, we have successfully implemented the CEMD, C(pH,E)MD and E-REMD methods in AMBER. These methods allow the theoretical study of systems at a constant value of redox potential, using both implicit and explicit solvent models.

Validation results and tests were presented for NAcMP8 axially connected to a histidine peptide. This system contains a single heme and three pH-active residues. Our simulations correctly describe the behavior of $E^o$ vs pH and of $pK_a$ vs redox potential, in good agreement with theoretical predictions. Regarding the NAcMP8's heme group, for the complete description of the pH dependence of its E° we conclude it is necessary to consider more than one pH-active group, as both propionates and also the glutamic acid interact significantly with the heme. We observed that the propionates have slightly different p$K_a$ values (a difference of ~0.15 pH units) and that these p$K_a$ values are in agreement with values obtained experimentally for other proteins that contain a single heme.

The addition of replica exchange significantly improves the statistical convergence of our E° predictions. The increase in the exchange attempt frequency in our replica exchange simulations did not show a significant change in the convergence efficiency of our calculations. The smallest EAF used (0.005ps$^{-1}$) was already satisfactory to reproduce effectively the same convergence as the larger EAF used.

Based on our computational benchmarks, we see that our methodologies have an acceptable computational cost in comparison to regular MD and that the GPU-accelerated code is able to provide high-performance in comparison to calculations using CPUs.

**AUTHOR INFORMATION**

**Corresponding Author**

\* E-mail: roitberg@ufl.edu

**Notes**

The authors declare no competing financial interest.

**ACKNOWLEDGMENTS**

The authors gratefully acknowledge financial support from CAPES (Brazil) and CNPq (Brazil; PDE fellowship for M.S.A., process number 234282/2014-2). This research is part of the Blue Waters sustained-petascale computing project, which is supported by the National Science Foundation (awards OCI-0725070 and ACI-1238993) and the state of Illinois. Blue Waters is a joint effort of the University of Illinois at Urbana-Champaign and its National Center for Supercomputing Applications. The authors acknowledge J. Swails, A. M. Baptista and M. R. Gunner for interesting discussions, and S. Lenka for a careful reading of the manuscript.